\documentclass[prl,
reprint,
 amsmath,amssymb,amscd
 aps,
]{revtex4-2}
\usepackage{mathrsfs}
\usepackage{graphicx}% Include figure files
\usepackage{dcolumn}% Align table columns on decimal point
\usepackage{bm}% bold math
\usepackage[dvipsnames,table,xcdraw]{xcolor}
\usepackage[
      colorlinks=true,
      linkcolor=black,  
      urlcolor=black,    
      filecolor=black,     
      citecolor=black,
      linktocpage=true,
      pdfstartview=FitV,
      bookmarksopen=true     
      ]{hyperref}
\usepackage{lmodern} 
\usepackage{amsmath}
\usepackage{amssymb}
\usepackage{graphicx,scalerel}

\allowdisplaybreaks

\begin{document}

\preprint{APS/123-QED}

\title{A Celestial Matrix Model}
\author{Arjun Kar}
\email{arjunkar@phas.ubc.ca}
\affiliation{%
Department of Physics and Astronomy \\
University of British Columbia \\
Vancouver, BC V6T 1Z1, Canada  
}%

\author{Lampros Lamprou}
\email{llamprou@phas.ubc.ca}
\affiliation{%
Department of Physics and Astronomy \\
University of British Columbia \\
Vancouver, BC V6T 1Z1, Canada  
}%

\author{Charles Marteau}
\email{cmarteau@phas.ubc.ca}
\affiliation{%
Department of Physics and Astronomy \\
University of British Columbia \\
Vancouver, BC V6T 1Z1, Canada  
}%

\author{Felipe Rosso}
\email{feliperosso@phas.ubc.ca}
\affiliation{%
Department of Physics and Astronomy \\
University of British Columbia \\
Vancouver, BC V6T 1Z1, Canada  
}%

% \date{\today}

\begin{abstract}
We construct a Hermitian random matrix model that provides a stable non-perturbative completion of Cangemi-Jackiw (CJ) gravity, a two-dimensional theory of flat spacetimes. The matrix model reproduces, to all orders in the topological expansion, the Euclidean partition function of CJ gravity with an arbitrary number of boundaries. The non-perturbative completion enables the exact computation of observables in flat space quantum gravity which we use to explicitly characterize the Bondi Hamiltonian spectrum.  We discuss the implications of our results for the flat space $S$-matrix and black holes.

\end{abstract}

\maketitle

%%%%%%%%%%%%%%%%%%%%%%%%%%%%%%%%%

\noindent \textbf{Introduction:} What is the mathematical description of a quantum Universe with vanishing cosmological constant? It has long been understood \cite{DeWitt:1967ub} that the gauge invariant content of a gravitational theory that is asymptotically flat is the map from initial states at $\mathscr{I}^-$ to final states at $\mathscr{I}^+$, i.e. an $S$-matrix. The path of the ambitious researcher attacking the problem of flat quantum gravity is thus, in principle, mapped out: Construct the relevant asymptotic Hilbert spaces and provide quantum mechanical rules for computing $S$-matrix elements, non-perturbatively. 

A surge of recent activity has outlined an interesting candidate formalism for this aspirational theory of flat Universes:
\emph{celestial holography}. Its central conceptual element is the translation of $S$-matrix elements in $d$-dimensional spacetime to correlation functions in a yet-unknown microscopic Euclidean quantum theory living on the $(d-2)$-dimensional celestial sphere. In four dimensions, this program has succeeded in elucidating the physics of soft graviton scattering by trading traditional soft theorems \cite{Weinberg:1965nx} for celestial Ward identities \cite{Strominger:2013jfa,He:2014laa}, providing tantalizing clues about the kinematic structure of this putative ``holographic'' dual. Its dynamical definition, however, remains elusive; what is missing is a prescription for generating the aforementioned correlators without reference to the bulk (see a recent attempt in \cite{Pasterski:2022joy}). In this work, we endeavour a first step towards rectifying this situation.

Encouraged by the wealth of insights low-dimensional versions of AdS/CFT holography have recently generated \cite{Maldacena:2016upp,Jensen:2016pah,Engelsoy:2016xyb,Saad:2019lba}, we consider the simplest non-trivial gravitational theory of two-dimensional flat spacetimes, introduced by Cangemi and Jackiw (CJ) in \cite{Cangemi:1992bj}. Its action reads  
\begin{align}\label{eq:CJaction}
I_{\rm CJ}=S_0\chi &+\frac{1}{2}\int_{\mathcal{M}} d^2x\sqrt{-g}
\big[ 
\Phi R+2\Psi(1-\varepsilon^{\mu \nu}\partial_\mu A_\nu)
\big] \nonumber \\[2pt] &+\frac{1}{2}\int_{\partial \mathcal{M}} du \sqrt{-h}\big[ 
2\Phi K -n^\mu \nabla_\mu \Phi
\big]\ ,
\end{align}
with $S_0\in \mathbb{R}_+$,  $\chi$ the Euler characteristic of the manifold $\mathcal{M}$, and $\varepsilon^{\mu \nu}$ the Levi-Civita tensor. The boundary terms are chosen to ensure the finiteness of the on-shell action, as well as the vanishing of its variations, when the latter obey certain asymptotic conditions. 

This is a dilaton-gravity theory constructed from the flat space limit of Jackiw-Teitelboim (JT) gravity \cite{Jackiw:1984je,Teitelboim:1983ux} by coupling it to a minimal matter sector: A topological abelian gauge field $A_\mu$ whose canonical momentum $\Psi$ endows the Universe with a ``vacuum'' energy. The special matter sector plays a central role in the quantum description of the theory and its non-perturbative properties. By ``freezing'' the momentum $\Psi$ of the gauge field,
the theory reduces to the dilaton-gravity sector of the widely known matterless CGHS model \cite{Callan:1992rs}. 
The purpose of this letter is to report progress towards a microscopic description of quantum CJ gravity.

At face value, the combined lessons of celestial holography \cite{deBoer:2003vf,Kapec:2014opa} and low-dimensional AdS/CFT \cite{Saad:2019lba} suggest that the non-perturbative description of CJ gravity involves an ensemble of zero-dimensional quantum theories, i.e. a \emph{celestial matrix model}. Taking this expectation seriously raises two fundamental questions: What is the ``holographic dictionary'' for converting gravitational questions to matrix model computations? And can we gain non-perturbative control of CJ gravity by identifying the precise dual matrix ensemble? In what follows, we summarize the answers to both questions, reserving technical details for a longer article \cite{ToAppear}.

\begin{figure}
    \centering
    \includegraphics[scale=0.32]{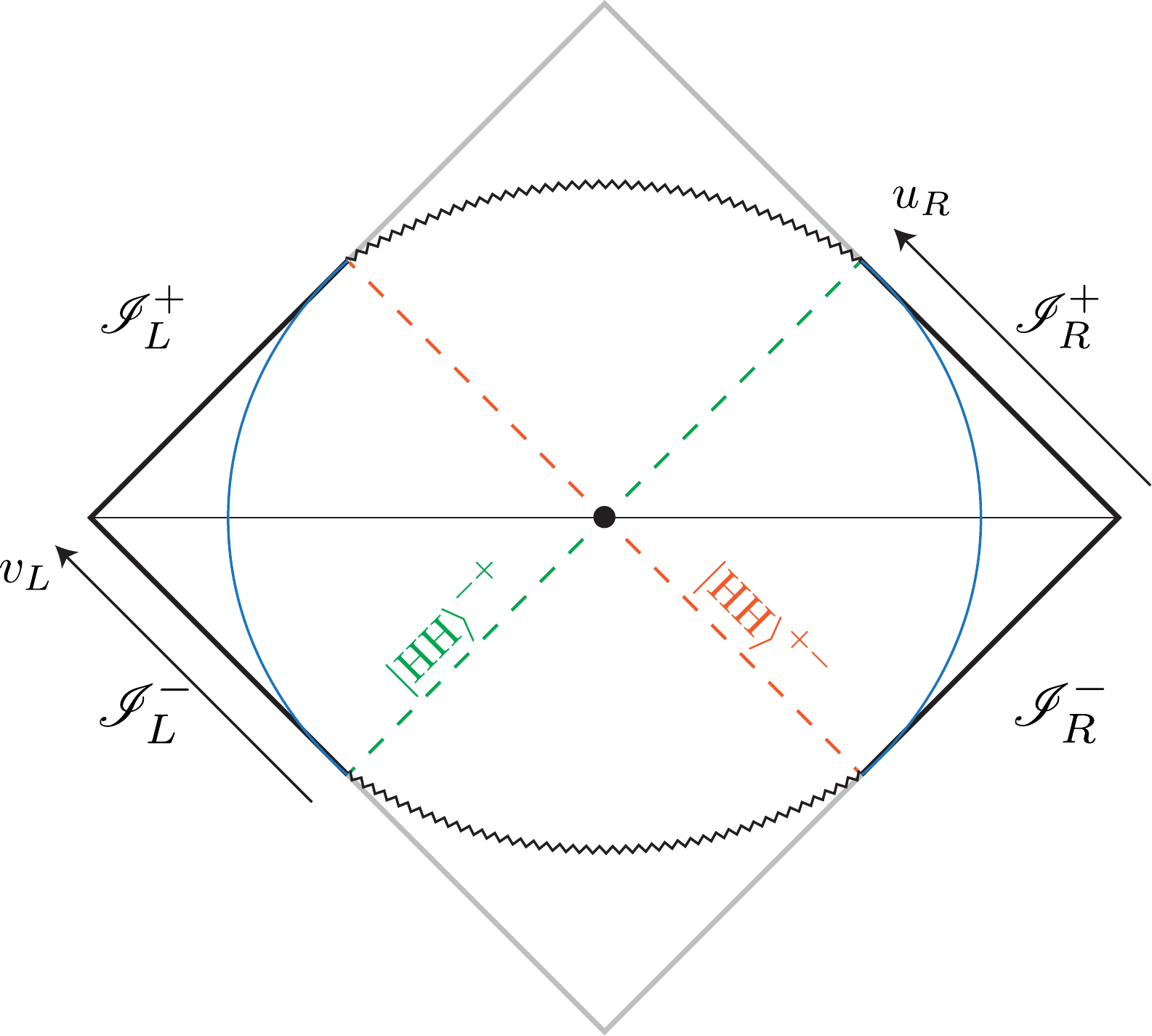}
    \caption{Penrose diagram of the Lorentzian solution. Two copies of \eqref{CGHSbondi} are needed: the right one involves the retarded time $u_R$ while the left one the advanced time $v_L$, so that together they cover the entire patch. The dilaton diverges positively at $\mathscr{I}^\pm_{R/L}$ while it takes some large negative value on the singularity.
    }\label{fig:Rindler}
\end{figure}

\noindent \textbf{Classical Cangemi-Jackiw Gravity:} All solutions of CJ gravity describe a rigid flat spacetime, filled with a homogeneous vacuum energy $\Psi(x^+,x^-)=\Lambda$ and permeated by a \emph{fixed} background electric flux $F=-dx^+\wedge dx^-$ and a dynamical dilaton field $\Phi(x^+,x^-)$. Its dynamics can be recast in gravitational language by expressing the solutions in the reference frame of the dilaton. Upon introducing a dilaton-Bondi coordinate frame $(r,u)$ via~\footnote{The following discussion is focused on describing the future null infinity and its relevant asymptotic states. The past null asymptotic are characterized in a directly analogous way, by introducing instead a ``past'' dilaton-Bondi frame $(r,v)$.}
\begin{equation}
    r=\frac{1}{\gamma}\,\Phi(x^+,x^-)\ , \qquad \frac{\partial}{\partial u} = \frac{1}{\gamma}\,\varepsilon^{\mu\nu} \partial_\nu \Phi(x^+,x^-)\,\partial_\mu\ , \label{dilatonframe}
\end{equation}
where $\gamma$ is a unit of inverse length introduced for dimensional consistency, the general solution consists of a pair of Rindler wedges in the configuration (Figure~\ref{fig:Rindler})
\begin{equation}\label{CGHSbondi}
\begin{aligned}
ds^2&= 
    -
    \frac{2\Lambda}{\gamma}
    \Big(r-\frac{\phi_h}{\gamma}\Big)du^2 - 2dudr \ ,
    \hspace{11 pt} \Phi= 
    \gamma r
    \ ,\\
    A&=-
    \Big(r-\frac{\phi_h}{\gamma}\Big)du + dg(r)\ , \hspace{33 pt} 
    \Psi=\Lambda\ ,
\end{aligned}
\end{equation}
where we also partially fixed the gauge for $A$. The vacuum energy $\Lambda$ controls the Rindler inverse temperature $\beta=2\pi \gamma/\Lambda$, while $\phi_h$ specifies the location of the bifurcate horizon. The dynamical variables that complete the phase space above are the difference of the gauge parameter $g(r)$ between the two asymptotic boundaries (the Wilson line of $A$) and the relative synchronization of the two asymptotic Bondi clocks (the ``gravitational Wilson line'').

Cangemi and Jackiw showed \cite{Cangemi:1992bj} that (\ref{eq:CJaction}) is equivalent to a BF theory for the Maxwell algebra
\begin{equation}\label{eq:Maxwell}
[P_a,K]=\epsilon_a^{\,\,\, b}P_b\ ,
\qquad 
[P_a,P_b]=\epsilon_{ab} Q\ ,
\end{equation}
with $\epsilon_{ab}$ the Levi-Civita symbol. Besides the translations $P_a$ and boost generator $K$ expected from the Poincar\'{e} symmetry of flat space, the algebra includes a non-trivial central extension $Q$ due to the presence of the matter sector in (\ref{eq:CJaction}). This fact renders it the minimal \textit{metric} Lie algebra containing $\mathfrak{iso}(1,1)$, which is crucial for a ``standard" definition of a BF theory \cite{Grumiller:2020elf}. 
As the BF description of low-dimensional AdS/CFT models has been fundamental in defining the measure of the Euclidean path integral \cite{Saad:2019lba}, this highlights the importance of considering the full CJ theory over the naive flat limit of JT gravity or the matterless CGHS model (see however \cite{Afshar:2021qvi}).

\noindent \textbf{Asymptotic Boundary Conditions:} In addition to the classical solutions of pure CJ gravity, it is important to characterize the allowed \emph{off-shell} configuration space by taming the fields' asymptotic behavior. This is important for the quantization of the theory, which amounts to specifying the functional domain of the path integration and a measure on it.

Starting from a reference classical solution (\ref{CGHSbondi}), e.g. the one with $(\Lambda,\phi_h)=(\gamma^2,0)$, a natural off-shell space includes all field configurations whose asymptotic expansion about $r=\infty$ is of \emph{the same form} as that of the parent solution. As we explain in \cite{ToAppear}, asymptotic fluctuations of the dilaton can always be absorbed into $g_{\mu \nu}$ and $A_\mu$ via a diffeomorphism that preserves their asymptotic form. With the additional requirement of vanishing on-shell action variations, we arrive at the asymptotic conditions
\begin{align}
    \Phi(r,u) &=\gamma r+ O\left(1/r\right)\  \label{dilatonbc}\\[4pt]
     \zeta^\mu A_\mu &=-r +O\left(1/r\right)\  \label{asymptoticgauge}\\[4pt]
    ds^2 &= -2(P(u) r +T(u) ) du^2 - 2dudr +O\left(1/r\right) \label{metricfalloff2}
\end{align}
with $\zeta^\mu =\frac{1}{\gamma}\varepsilon^{\mu\nu}\partial_\nu \Phi$ an asymptotic timelike vector that generalizes (\ref{dilatonframe}), and $(P(u),T(u))$ arbitrary smooth functions that characterize the future asymptotic states. 

Condition (\ref{dilatonbc}) is key. The frozen dilaton profile near null infinity allows us to define a regulated \emph{asymptotic boundary} of the spacetime $\partial{\cal M}_\epsilon$ and a \emph{canonical time flow} along it for all off-shell configurations via 
\begin{equation}
    \Phi\big|_{\partial {\cal M}_\epsilon} = \frac{1}{\epsilon}\, , 
    \qquad \quad \frac{\partial}{\partial u} = \frac{1}{\gamma} \varepsilon^{\mu\nu} \partial_\nu \Phi \big|_{\partial {\cal M}_\epsilon} \partial_\mu\ , \label{boundarydef}
\end{equation}
respectively, since neither the value nor the normal derivative of $\Phi$ fluctuate in the $\epsilon \to 0$ limit. On-shell, (\ref{boundarydef}) selects a constant acceleration cutoff in each Rindler wedge (Figure \ref{fig:Rindler}) and the limits $\lim_{\epsilon\to 0}{\partial {\cal M}}_{R,\epsilon} = \mathscr{I}_R^+$, $\lim_{\epsilon\to 0}{\partial {\cal M}}_{L,\epsilon} = \mathscr{I}_L^-$ define right future and left past null infinity, respectively \footnote{The reason our limit gives $\mathscr{I}^-_L$  and $\mathscr{I}_R^+$ instead of future null infinity in both wedges is that we want to send $r\to \infty$ while preserving the synchronization of the two Rindler clocks, i.e. without exciting the gravitational Wilson line. The only way to do this is to approach opposite sides of null infinity in the two wedges \cite{ToAppear}.} Condition (\ref{asymptoticgauge}) then fixes the pullback of $A_\mu$ on $\partial {\cal M}_\epsilon$.

In contrast, the asymptotic metric fluctuates freely. The class of metrics (\ref{metricfalloff2}) forms the coadjoint orbit of an asymptotic symmetry group which, in view of the constraint (\ref{asymptoticgauge}), is the warped Virasoro group \cite{Afshar:2019axx}.  All off-shell configurations are obtained from the parent ${(\Lambda,\phi_h)=(\gamma^2,0)}$ solution via a transformation
\begin{equation}
    u \to f(u)\, \, , \quad 
    r\to \frac{r+g'(u)}{f'(u)} \,\, , \quad 
    A \to A+dg(u)\ . \label{warpedVir}
\end{equation}
%\CM{The action of the group should be understood as acting only on A and g? Also, the subgroup take you out of the fall-offs (5,6,7)?}The orbit of the subgroup $(f(u),g(u))=(\frac{\Lambda }{\gamma^2},-\frac{\phi_h}{\gamma})u$  defines the on-shell subspace (\ref{CGHSbondi}). \textcolor{ForestGreen}{[FR] I changed the sign in $g(u)=-\frac{\phi_h}{\gamma}u$, erase this comment if you agree.} 
Asymptotic states of CJ gravity are, therefore, mapped to profiles of the frame variables $(f(u),g(u))$ on each component $\mathscr{I}^{\pm}_{L,R}$.

The fact that the boundary induced metric is not fixed by our boundary conditions is an important deviation from the standard Dirichlet condition in AdS$_2$ JT gravity.
The proper time along the asymptotic boundary is not a good clock in quantum CJ gravity, essentially because there is no notion of proper time that survives as $\epsilon\to 0$ and $\partial \mathcal{M}_\epsilon$ approaches the null surface $\mathscr{I}_R^+\cup \mathscr{I}_L^-$.
Instead, the holographic clock is defined by the dilaton field, with a rate set by its non-fluctuating normal derivative as in (\ref{boundarydef}). The treatment of the other asymptotic null infinity $\mathscr{I}_R^-\cup  \mathscr{I}_L^+$ is performed in a completely analogous way.

\noindent \textbf{The Holographic Dictionary:} The classical asymptotic states described above form a vector space quantum mechanically. Completing it into a Hilbert space is a key first step in defining the $S$-matrix of CJ gravity. To do so we must choose an inner product, a natural choice for which is provided by the Euclidean path integral 
\begin{equation}\label{eq:CJpartfunc}
Z(\beta)=\int 
\mathcal{D}g_{\mu \nu}\mathcal{D}A_\mu
\mathcal{D}\Phi \mathcal{D}\Psi  e^{-I_{\rm CJ}}\ ,
\end{equation}
where the boundary time $u$ defined via (\ref{boundarydef}) has been analytically continued to $u\to i \tau$ and compactified $\tau \sim \tau +\beta$ by enforcing periodic boundary conditions. Expression (\ref{eq:CJpartfunc}) can be interpreted in two ways: (a) As the ``partition function'' of CJ gravity, probing the spectrum of the Hamiltonian generator of Bondi time along $\mathscr{I}_R^+$, or (b) as the overlap of two Hartle-Hawking (HH) states on the Hilbert space ${\cal H}_{\mathscr{I}^-_L\cup\mathscr{I}^+_R}\equiv \mathcal{H}^{-+}$ produced by each half of the Euclidean circle. This state should be thought of as living on the horizon that connects the two asymptotic regions (Figure~\ref{fig:Rindler}). Decorating the partition function with operator insertions, then, endows the asymptotic Hilbert space $\mathcal{H}^{-+}$ with an inner product. Equivalently, from the analytic continuation of the right advanced time, one can compute the trace of the corresponding Hamiltonian on $\mathscr{I}^-_R$ and construct ${\cal H}_{\mathscr{I}^+_L\cup\mathscr{I}^-_R}\equiv \mathcal{H}^{+-}$.

The path integral (\ref{eq:CJpartfunc}) should be understood mathematically as a sum over connected manifolds with different topologies and a path integration over field configurations on each of them. The domain of the latter is the off-shell functional space (\ref{dilatonbc}--\ref{metricfalloff2}). The measure is, in turn, obtained rigorously from the BF description of the theory \cite{ToAppear}. The partition function (\ref{eq:CJpartfunc}) can further be generalized to a Euclidean path integral $Z(\beta_1,\dots,\beta_n)$ with an arbitrary number $n$ of asymptotic boundaries (\ref{boundarydef}) with periodicities $\beta_i$, $i=1,\dots n$. Non-factorized contributions to $Z(\beta_1,\dots,\beta_n)$ may be interpreted as computing expectation values of inner products over an ensemble of microscopic theories \cite{Saad:2019lba}. 

Our proposed holographic dictionary can now be articulated by closely following the steps of the successful AdS$_2$ case \cite{Saad:2019lba}. Our claim is that the $n$-boundary partition function is non-perturbatively computed by the $n$-point function of a single-trace matrix operator $\mathbb{O}(\beta)$ in a random matrix ensemble
\begin{equation}\label{dictionary}
Z(\beta_1,\dots,\beta_n)= \langle
\prod_{i=1}^n\mathbb{O}(\beta_i)\rangle_c\ ,
\end{equation}
where the subscript $c$ means the connected ensemble average. We shall establish (\ref{dictionary}) by explicitly matching both sides of this equality to all orders in perturbation theory. This is the task we turn to next.

\noindent \textbf{Euclidean Partition Function:} The gravitational expression (\ref{eq:CJpartfunc}) can be studied in a series expansion in $e^{-S_0}$, called the ``topological expansion''. Since the Euler characteristic of a two-dimensional manifold is ${\chi=2(1-g)-n}$, with $g$ and $n$ the genus and number of boundaries, the topological expansion reads
\begin{equation}\label{eq:topological}
Z(\beta_1,\dots,\beta_n)\simeq 
\sum_{g=0}^{\infty}
(e^{-S_0})^{2(g-1)+n}
Z_g(\beta_1,\dots,\beta_n)
\end{equation}
where the symbol $\simeq$ highlights the fact the equality holds up to non-perturbative corrections in $e^{-S_0}$. The quantity $Z_g(\beta_1,\dots,\beta_n)$ is defined as (\ref{eq:CJpartfunc}), with the important difference the path integral over metrics is constrained to manifolds of fixed genus $g$.

Due to the linear dependence of (\ref{eq:CJaction}) on $\Phi$, after a rotation of its contour along the imaginary axis, the dilaton integration yields a Dirac delta $\delta(R)$ which forces all off-shell metrics to be flat. As explained in \cite{Godet:2021cdl}, there are no locally flat manifolds with more than two asymptotic boundaries \footnote{In this context, an asymptotic boundary is defined as one for which the distance of any bulk point to the boundary is infinite. This excludes geometries that would contribute to (\ref{eq:vanishing}), like a flat plane with an arbitrary number of holes.}. Hence, the entire topological expansion (\ref{eq:topological}) vanishes for
    \begin{equation}\label{eq:vanishing}
    Z(\beta_1,\dots,\beta_n)\simeq 0 \  , \qquad  n\ge 3\ .
    \end{equation}
When $n=1,2$, only the $g=0$ terms in (\ref{eq:topological}) are non-zero, corresponding to the disk and cylinder topologies. The path integrals in these two cases reduce to the integral over a quantum mechanical system \cite{Afshar:2019axx} and are one-loop exact \cite{Stanford:2017thb,Afshar:2019tvp}, and can therefore be computed exactly \cite{Afshar:2019tvp,Godet:2020xpk} with respect to the measure derived from the bulk BF description of CJ gravity \cite{ToAppear}. In a particular normalization for the measure, one finds
\begin{equation}\label{eq:CJdiskcyl}
\begin{aligned}
Z(\beta) &  \simeq e^{S_0}Z_{\rm disk}(\beta)=e^{S_0}
\frac{\pi^3}{2(\gamma \beta)^2}\ ,\\[3pt]
Z(\beta_1,\beta_2) & \simeq Z_{\rm cylinder}(\beta_1,\beta_2)=
\frac{1}{\gamma(\beta_1+\beta_2)} \ .
\end{aligned}
\end{equation}
The spectral density is $\varrho(E)\simeq e^{S_0}\frac{\pi^3}{2\gamma^2}E$, obtained from the inverse Laplace transform of $Z(\beta)$.

%%%%%%%%%%%%%%%%%%%%%%%%%%%%%%%%%
\noindent \textbf{Celestial Matrix Model:} Completing the holographic dictionary (\ref{dictionary}) amounts to specifying the following matrix model data:
\begin{itemize}
\setlength\itemsep{0.03em}
    \item[a.] Symmetry class of the matrix $M$.
    \item[b.] Probability measure over the ensemble.
    \item[c.] Matrix operator $\mathbb{O}(\beta)$ corresponding to the insertion of a boundary in the gravity partition function.
\end{itemize}
The criterion of success is the ability of the relevant ensemble expectation values to reproduce all orders of the topological expansion of CJ gravity (\ref{eq:vanishing}) and (\ref{eq:CJdiskcyl}).
Interestingly, there is a unique choice that achieves this matching.

There exist ten symmetry classes of random matrix models \cite{Dyson:1962es,1997,Stanford:2019vob}. As explained in \cite{Stanford:2019vob}, for path integrals over orientable surfaces, the $N$-dimensional square matrix $M$ must either be arbitrary complex or Hermitian. The structure of the model is, then, studied in $1/N$ perturbation theory, via the loop equations \cite{Eynard:2004mh,Stanford:2019vob}. These are recursion relations dependent only on the model's symmetry class and a complex function $y(z)$, the spectral curve, encoding the leading average spectral density of the ensemble. Condition (\ref{eq:vanishing}) fixes the analytic structure of $y(z)$, placing a branch cut along the whole real line \footnote{See sections 4.1 and 5.2.1 of \cite{Stanford:2019vob} for a more detailed discussion on how this cancellation comes about.}. The specific choice of $y(z)$ and operator $\mathbb{O}(\beta)$ can then be deduced from the disk and cylinder partition functions (\ref{eq:CJdiskcyl}).

Our celestial matrix model is defined as follows. Start with an ensemble of finite $N$ Hermitian matrices $M$ with probability measure $dMe^{-N\,{\rm Tr}V(M)}$  \footnote{In the double scaling limit, the details of the potential $V(M)$ away from the origin are non-universal and therefore irrelevant. There is an infinite class of potentials that result in the same double scaled model \cite{Claeys}.}
\begin{equation}\label{eq:MatrixPotential}
V(M)=M^4-\frac{1}{2}M^2\ ,
\end{equation}
which for large $N$ gives the leading eigenvalue density ${\rho_0(\lambda)=\frac{\lambda^2}{2\pi}\sqrt{4-\lambda^2}}$. The spectral curve $y(z)$ is obtained from ${y(\lambda \pm i\epsilon)=\mp i\pi \rho_0(\lambda)}$. For $y(z)$ to have a branch cut along the whole real line one needs to take a double scaling limit \cite{Ginsparg:1993is,DiFrancesco:1993cyw}, which amounts taking $N$ large while simultaneously rescaling the eigenvalues $\lambda_i$ of $M$ near the origin, where $\rho_0(\lambda)\sim \lambda^2$. In this limit, the matrix model parameters $(N,\lambda_i)$ are replaced by $(\hbar,\alpha_i)$, according to $1/N=\hbar(\gamma^{3/2}/\pi^3)\delta^3$ and $\lambda_i=\alpha_i\delta$,
where double scaling corresponds to $\delta\rightarrow 0$. This simple model has been extensively studied in other contexts \cite{Douglas:1990xv,Crnkovic:1990mr,10.2307/121101,Bleher:2002ys}. The $1/N$ expansion becomes an expansion in $\hbar$, where we shall identify ${\hbar=e^{-S_0}}$ to make contact with gravity (\ref{eq:topological}). The eigenvalue density to leading order in $\hbar$ in the double scaling limit $\rho_0(\alpha)$ becomes
\begin{equation}
\rho_0(\alpha)=
\lim_{\delta,\hbar \rightarrow 0}\,
\langle {\rm Tr}\,\delta(\bar{M}-\alpha) \rangle=
\frac{\pi^2}{\hbar \gamma^{3/2}}\alpha^2 \ ,
\end{equation}
where $\bar{M}$ is the random matrix with the rescaled eigenvalues $\alpha_i$. The spectral curve $y(z)$ associated to $\rho_0(\alpha)$ has a branch cut along the whole real line, ensuring through the loop equations the vanishing of all perturbative corrections to trace class observables, except for the leading single and double trace cases. All that is left to do is pick the operator $\mathbb{O}(\beta)$ that ensures the matching with (\ref{eq:CJdiskcyl}), which gives
\begin{equation}\label{eq:Operator}
\mathbb{O}(\beta)=
\int_{-\infty}^{+\infty}
\frac{dp}{\sqrt{\gamma}}
\,{\rm Tr}\,e^{-\beta(\bar{M}^2+p^2)}\ ,
\end{equation}
where the integral over $p$ is unaffected by the ensemble average over $\bar{M}$. Quite remarkably, $\mathbb{O}(\beta)$ has the form one would expect for the trace of $e^{-\beta H}$ for some Hamiltonian $H$, which contains a contribution from the random matrix $\bar{M}$ and the momentum $p\in \mathbb{R}$ of a free particle.

The double scaled Hermitian matrix model with probability measure defined from (\ref{eq:MatrixPotential}) reproduces the topological expansion of CJ gravity to all orders. This result is highly non-trivial. The underlying relation between the disk and cylinder partition functions was crucial for the existence of a solution to our problem. For instance, if $Z_{\rm disk}(\beta)$ had a half-integer power of $\beta$ instead of even, one could show there is no matrix model that reproduces the gravitational results with the same $Z_{\rm cylinder}(\beta_1,\beta_2)$, for any operator $\mathbb{O}(\beta)$.

The peculiar integration over the continuous variable $p$ appearing in the matrix operator $\mathbb{O}(\beta)$ arises due to the additional generator in the Maxwell algebra \eqref{eq:Maxwell} which was needed to obtain a BF description of CJ gravity.
The fact that it leads to a continuous spectrum for $\mathbb{O}(\beta)$, which is a departure from the situation in JT gravity \cite{Saad:2019lba}, may be related to the fact that CJ gravity describes flat space which must be thought of as a quantum system in infinite volume rather than a covariant AdS ``box''.

\begin{figure}
    \centering
    \includegraphics[scale=0.42]{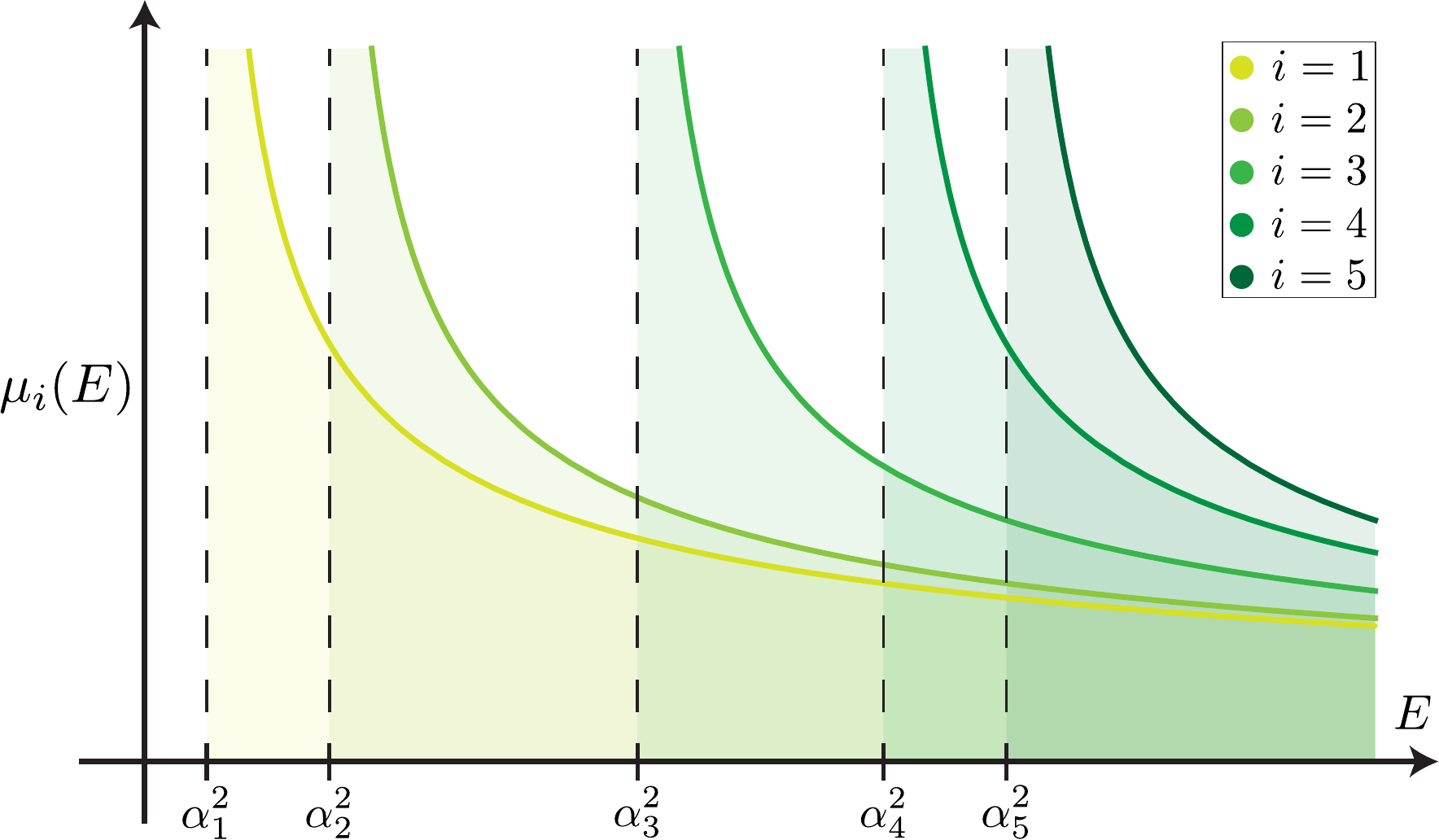}
    \caption{Plot of each of the discrete contributions $\mu_i(E)$ in (\ref{eq:CJNonpert}) for a single instance of the ensemble of CJ gravity. The analogous diagram for JT gravity \cite{Johnson:2022wsr} corresponds to a discrete collection of Dirac deltas. These curves are smoothened when averaging over the ensemble $\langle \mu_i(E) \rangle$, as shown by the green curves in Figure \ref{fig:rhoCJ}.}\label{fig:singleCJ}
\end{figure}

%%%%%%%%%%%%%%%%%%%%%%%%%%%%%%%%%%
\noindent \textbf{Non-Perturbative Bondi Spectrum:} The Celestial Matrix Model can be now used to explicitly characterize non-perturbatively the spectral density $\varrho(E)$ of the Hamiltonian generator of Bondi time along null infinity. Using (\ref{dictionary}) and the fact $\varrho(E)$ is defined from the inverse Laplace transform of $Z(\beta)$, one finds
\begin{equation}\label{eq:CJNonpert}
\varrho(E)=
\frac{2}{\sqrt{\gamma}}
\sum_{i=1}^{\infty}
\Big\langle 
\frac{\Theta(E-\alpha_i^2)}{\sqrt{\smash[b]{E-\alpha_i^2}}}
\Big\rangle \equiv \sum_{i=1}^{\infty}\langle \mu_i(E) \rangle\ ,
\end{equation}
where we have defined $\mu_i(E)$. Given the matrix operator (\ref{eq:Operator}) and noting $1/\sqrt{E}$ corresponds to the spectral density of a free particle, the structure of $\varrho(E)$ is quite intuitive: it corresponds to a superposition of free particle densities centered at random positions $\alpha_i^2$ determined by the matrix $\bar{M}$. In Figure~\ref{fig:singleCJ} we plot each of the discrete contributions $\mu_i(E)$ appearing in (\ref{eq:CJNonpert}) for a single instance of CJ gravity (i.e. without the ensemble average) so that the actual spectrum is obtained by summing these curves. The analogous quantity in JT gravity is $\mu_i^{\rm JT}(E)=\delta(E-\alpha_i)$.

Let us now compute the full spectral density $\varrho(E)$ and each of the individual terms $\langle \mu_i(E) \rangle$ in (\ref{eq:CJNonpert}), including non-perturbative effects. To do so, we must study the double scaled matrix model using the method of orthogonal polynomials \cite{Ginsparg:1993is,DiFrancesco:1993cyw,Eynard:2015aea} (developed for the JT gravity case in \cite{Johnson:2019eik,Johnson:2020exp,Okuyama:2019xbv}) that is well suited for computing quantities beyond perturbation theory. All observables are ultimately determined by the matrix model kernel $K(\alpha,\bar{\alpha})$, for instance the eigenvalue spectral density is obtained from its diagonal components $\bar{\alpha}=\alpha$. 

The recipe for computing $K(\alpha,\bar{\alpha})$ including non-perturbative contributions proceeds as follows \footnote{For the interested reader, the details of this procedure, including an explicit derivation of this formalism for the celestial matrix model will appear in \cite{ToAppear}.}. After picking specific values for $(S_0,\gamma)$ one numerically solves the string equation \cite{Douglas:1990xv}
\begin{equation}\label{eq:StringEq}
\frac{\pi^3}{2\gamma^{3/2}}
\Big[
r(x)^3-\frac{1}{2}\hbar^2r''(x)
\Big]+r(x)x=0\ ,
\end{equation}
and obtains $r(x)$. Constructing the Schrodinger operators $\mathcal{H}_{s}=-\hbar^2\partial_x^2+[r(x)^2-s\hbar r'(x)]$ with $s=\pm 1$, one numerically computes their eigenfunctions $\varphi_s(x,\alpha)$ with eigenvalue $\alpha^2$. Finally, the kernel $K(\alpha,\bar{\alpha})$ is obtained by combining these eigenfunctions as~\cite{Rosso:2021orf}
\begin{equation}\label{eq:Kernel}
K(\alpha,\bar{\alpha})=
\sum_{s=\pm}\int_{-\infty}^0dx\,\varphi_s(x,\alpha)\varphi_s(x,\bar{\alpha})\ .
\end{equation}

\begin{figure}
    \centering
    \includegraphics[scale=0.58]{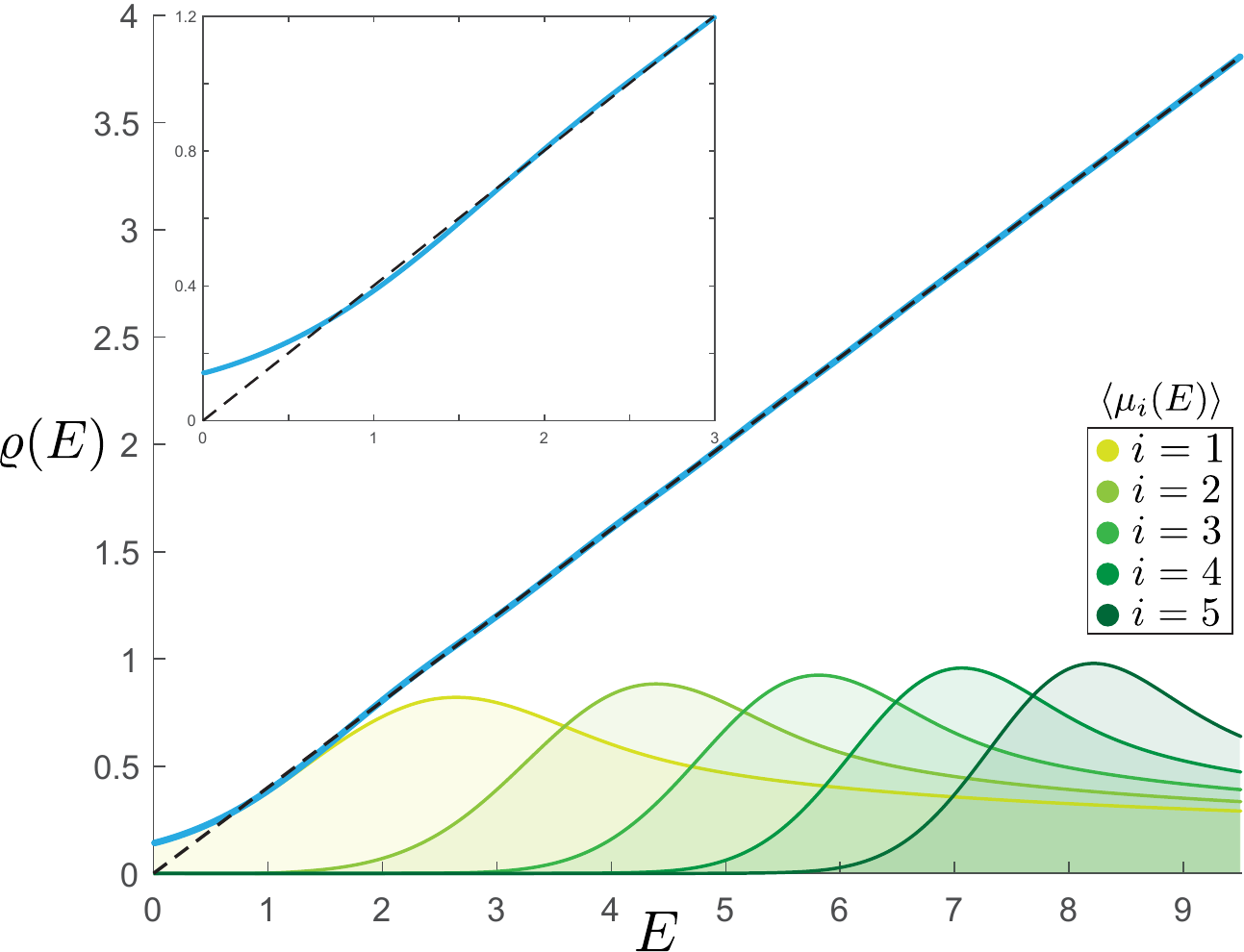}
    \caption{Spectrum of CJ gravity with $(S_0,\gamma)=(0,\pi^2/2^{2/3})$. The solid blue line gives the full non-perturbative spectral density $\varrho(E)$ which oscillates around the perturbative result
    $\varrho(E)\simeq e^{S_0}\frac{\pi^3}{2\gamma^2}E$ obtained from (\ref{eq:CJdiskcyl}). The green curves correspond to the first five individual contributions $\langle \mu_i(E) \rangle$ in the infinite sum in (\ref{eq:CJNonpert}), computed using the methods of \cite{Bornemann_2009,Johnson:2022wsr,Johnson:2021zuo}.}\label{fig:rhoCJ}
\end{figure}

Applying this procedure, we obtain the full eigenvalue spectral density from the diagonal components of $K(\alpha,\bar{\alpha})$, that is used in (\ref{eq:CJNonpert}) to compute $\varrho(E)$. The solid blue line in Figure \ref{fig:rhoCJ} shows the final result for $\varrho(E)$, where non-perturbative effects generate oscillations around the dashed line, corresponding to the simple linear answer $\varrho(E)\simeq e^{S_0}\frac{\pi^3}{2\gamma^2}E$ valid to all orders in perturbation theory. While non-perturbative effects are suppressed at high energies, they dominate the low energy behavior of the spectrum, particularly at zero energy, where we find a non-zero density of states.

One can take this formalism further and use $K(\alpha,\bar{\alpha})$ to extract fine grained information about the spectrum of CJ gravity, like the individual terms $\langle \mu_i(E) \rangle$ appearing in (\ref{eq:CJNonpert}). To do so, one needs the probability density function associated to each individual eigenvalue $\alpha_i$ of the matrix $\bar{M}$, which can be computed from an appropriately defined Fredholm determinant using the numerical methods developed in \cite{Bornemann_2009,Johnson:2022wsr,Johnson:2021zuo}. Applying this procedure (see \cite{ToAppear} for details), we compute $\langle \mu_i(E) \rangle$, with the $i=1,\dots,5$ cases shown in Figure~\ref{fig:rhoCJ}. Comparing with the spectrum of a single instance of CJ gravity in Figure~\ref{fig:singleCJ}, the green curves of Figure~\ref{fig:rhoCJ} show how $\mu_i(E)$ is smoothened when averaged over the ensemble.

\noindent \textbf{The $S$-Matrix and Higher Dimensions:}
Using the celestial matrix model, we can try to construct a non-perturbative $S$-matrix for CJ gravity.
Following the calculation of correlation functions in JT gravity \cite{Yang:2018gdb,Saad:2019pqd,Iliesiu:2021ari}, we may define a perturbative $S$-matrix \cite{Dray:1984ha,tHooft:1996rdg} as a bulk operator mapping $\mathcal{H}^{+-}$ to $\mathcal{H}^{-+}$.
Non-perturbative $S$-matrix elements $\langle \text{out} | S |\text{in}\rangle$ are then obtained by writing the states using the Euclidean path integral, possibly with probe operator insertions, and evaluating them using the celestial matrix model.
The perturbative $S$-matrix and any probe operator insertions are to be treated as smooth functions of the Bondi energy for which a single matrix in the celestial ensemble provides a specific spectrum (Figure~\ref{fig:singleCJ}).

This also suggests a construction procedure for a pair of entangled black holes in any celestial CFT. 
While $\mathcal{H}^{+-}$ and $\mathcal{H}^{-+}$ do not factorize in two dimensions, we might expect that they do in higher dimensions as in AdS/CFT.
If so, our results suggest a pair of entangled black holes in flat space is associated to a thermofield double entangled state between e.g. $\mathcal{H}_R^+$ and $\mathcal{H}_L^-$ in a left and right celestial CFT, since this is the state created by the Hartle-Hawking procedure in higher dimensions.
While the entanglement structure is simple in the thermofield double state, to determine the incoming or outgoing scattering states we must act with either $S_L$ or $S_R$, the $S$-matrices of the left or right celestial CFTs, thereby complicating the entanglement structure \cite{tHooft:2016qoo,Betzios:2016yaq}.
We leave further exploration of the $S$-matrix of CJ gravity and non-perturbative formulation of black hole states in celestial CFTs for our longer article \cite{ToAppear} and future work.

%%%%%%%%%%%%%%%%%%%%%%%%%%%%%%%%%%
\noindent \textbf{Acknowledgments:} We thank Panos Betzios, Laura Donnay, Victor Godet, Clifford Johnson, Ana-Maria Raclariu and Romain Ruzziconi for discussions.  AK and LL are supported by the Simons Foundation via the It from Qubit Collaboration. CM and FR acknowledge support from NSERC. FR is also supported in part by the Simons Foundation.

\bibliography{References}
% \bibliography{apssamp}% Produces the bibliography via BibTeX.

\end{document}